\title{Yang-66P Paper}

\documentclass{aa}

\usepackage{graphicx}
\usepackage{txfonts}
\usepackage{xcolor}
\usepackage{subfig}
\usepackage{makecell}
\usepackage{color}
\begin{document} 

   \title{Comet 66P/du Toit: not a near Earth main belt comet}


  \author{
Bin Yang\inst{1}\and 
Emmanu{\"e}l Jehin \inst{2}\and 
Francisco J. Pozuelos \inst{2,4}\and
Youssef Moulane \inst{1,2,3}\and 
Yoshiharu Shinnaka \inst{5}\and 
Cyrielle Opitom \inst{1}\and 
Henry H. Hsieh \inst{6}\and 
Damien Hutsem{\'e}kers \inst{2}\and 
Jean Manfroid \inst{2}
}

\institute{
  European Southern Observatory, Alonso de C\`{o}rdova 3107,  Vitacura, Casilla 19001, Santiago, Chile \and
  Space sciences, Technologies \& Astrophysics Research (STAR) Institute,
  Universit\'e de Li\`ege, 4000 Li\`ege, Belgium \and
  Oukaimeden Observatory, High Energy Physics and Astrophysics Laboratory, Cadi Ayyad University, Marrakech, Morocco \and 
  EXOTIC Lab, UR Astrobiology, AGO Department, University of Li\`ege, 4000 Li\`ege, Belgium \and
  Laboratory of Infrared High-resolution Spectroscopy, Koyama Astronomical Observatory, Kyoto Sangyo University, Motoyama, Kamigamo, Kita-ku, Kyoto 603-8555, Japan \and 
  Planetary Science Institute, 1700 East Fort Lowell, Suite 106, Tucson, AZ 85719, USA
}


 
  \abstract
   {Main belt comets (MBCs) 
are a peculiar class of volatile-containing objects with comet-like morphology and asteroid-like orbits. However, MBCs are challenging targets to study remotely due to their small sizes and the relatively large distance they are from the Sun and the Earth. Recently, a number of weakly active short-period comets have been identified that might originate in the asteroid main belt. Among all of the known candidates, comet 66P/du Toit has been suggested to have one of the highest probabilities of coming from the main belt.}
   {The main goal of this study is to investigate the physical properties of 66P via spectroscopic and imaging observations to constrain its formation conditions. In particular, the isotopic abundance ratio and the ortho-to-para ratio (OPR) of gaseous species can be derived via high-resolution spectroscopy, which is sensitive to the formation temperature of the nucleus.}
   {We obtained medium and high-resolution spectra of 66P from 300-2500 nm with the X-shooter and the UVES instruments at the Very Large Telescope in July 2018. We also obtained a series of narrow-band images of 66P to monitor the gas and dust activity between May and July 2018 with TRAPPIST-South. In addition, we applied a dust model to characterize the dust coma of 66P and performed dynamical simulations to study the orbital evolution of 66P.}
   {We derive the OPR of ammonia (NH$_3$) in 66P to be 1.08$\pm$0.06, which corresponds to a nuclear spin temperature of $\sim$34 K. We computed the production rates of OH, NH, CN, C$_3,$ and C$_2$ radicals and measured the dust proxy, Af$\rho$. The dust analysis reveals that the coma can be best-fit with an anisotropic model and the peak dust production rate is about 55 kg s$^{-1}$ at the perihelion distance of 1.29 au. Dynamical simulations show that 66P is moderately asteroidal with the capture time, t$_{cap} \sim 10^4$ yr. 
   }
   {
  Our observations demonstrate that the measured physical properties of 66P are consistent with other typical short-period comets and differ significantly from other MBCs. Therefore, 66P is unlikely to have a main belt origin.
   }

   \maketitle
%

\section{Introduction}
Water is essential for life and it is an important tracer of the formation and evolution processes in planetary systems. In addition, the distribution of water and volatiles in our solar system is a primary determinant of habitability. Recent space mission results and ground-based observations show that water is prevalent 
throughout the solar system.\ This includes many previously unexpected locations, such as the asteroid main belt, where most known asteroids reside in the region between the orbits of Mars and Jupiter. The so called main belt comets (hereafter MBCs) are a peculiar population with a comet-like morphology and asteroid-like orbits, which most likely contain buried water ice \citep{Hsieh2006}. The MBCs, thus, are particularly important in relation to the history of water and other major volatiles. They are appealing targets for future space missions since in situ sampling could be possible with similar instruments as those on Rosetta \citep{Snodgrass2017bb}. However, MBCs are observationally challenging targets not only because they are small (merely a few km across), but also because they exhibit very low activity. Even with the most powerful telescopes (i.e., Herschel, Keck, VLT, GEMINI, and GTC), all of the previous attempts failed to detect any evidence of gaseous products of sublimation \citep{Hsieh2011,de2012upper,Licandro2013,Snodgrass2017}. These studies demonstrate that spectroscopic detection of gas at main-belt distances is extremely difficult due to the rapidly declining water sublimation rates from 2 au to 3 au \citep{Jewitt2015}. 

Recently, \cite{fernandez2015} identified a number of near-Earth short-period comets or Jupiter family comets (JFCs) that are more dynamically stable and exhibit weaker activity than other JFCs.\ Furthermore, they suggest that these objects might originate in the main asteroid belt. This possibility is also supported by \cite{Hsieh2016}, who find that main-belt asteroids can indeed attain JFC-type orbits under certain circumstances. We call these anomalous objects ``Near-Earth MBC candidates," or NEMBC candidates. Among all of the known NEMBC candidates, comet 66P is defined by the authors as highly asteroidal (while still satisfying the dynamical requirements to be considered a short-period comet), meaning that there is a high probability that this object comes from the main belt \citep{fernandez2015}. Thus, 66P is an attractive target because it may represent a sample of MBCs that is more accessible than their main belt counterparts. If their main belt origin is confirmed, the newly identified NEMBC candidates (such as 66P) represent the best sample with which we can hope to take a close look at MBCs and investigate their composition in details other than a spacecraft visit. In turn, studying NEMBCs will give us a window to estimate MBC compositions and test terrestrial water origin models.
If 66P forms in the asteroid belt, we would expect its physical properties, such as composition, isotopic ratio, and water abundance, to be distinct from those of typical JFCs. These properties can be derived through spectroscopic and photometric observations. In Section 2, we present observations performed with the ESO's Very Large Telescope (VLT) and the TRAPPIST-South telescope. In Section 3, we present the observational results as well as the results of our dust models. In Section 4, we revisit the dynamical evolution of 66P, following the approach in \cite{fernandez2015} but by using the latest orbital parameters. Lastly, in Section 5, we summarize all the observational and theoretical simulation results and present our conclusion on the origin of 66P.

\section{Observations and data reduction}  

\subsection{X-shooter observations}
The chemical composition of comets can be investigated by observing gas emission bands in the UV and visible, and potentially by observing absorption features in the near-infrared (NIR) continuum due to ice grains or various minerals in grains. X-shooter allows all possibilities to be explored in one exposure. Notably, the OH emission band at 308 nm can be observed in the bluest order of the UVB arm, which in turn is used to estimate the total water production rate. Traditionally, other gas production rates are compared to the water production rate as a reference.
We were allotted 2.7 hours of director discretionary time to observe 66P with X-shooter. The two sets of observations took place in service mode a few days apart in July 2018. Details on the observing geometry are given in Table \ref{rate66P}. We used slit widths 1.3$^{\prime\prime}$, 1.2$^{\prime\prime}$ , and 1.2$^{\prime\prime}$ in the UVB, VIS, and NIR arms, respectively. The spatial scale covers from the optocenter up to 3700 km in both directions. The overall wavelength range runs from 300-2500 nm with the spectral resolution of 4100 in the UVB arm, 6500 in the VIS arm, and 4300 in the NIR. We also observed the solar analog star SA93-101 immediately after the comet on both nights.

The X-SHOOTER data is reduced using the Reflex environment \citep{Freudling2013} based on the ESO XSHOOTER pipeline \citep{Modigliani2010}. We used the Reflex pipeline to obtain the two dimensional order-merged and wavelength-calibrated spectra, while the one-dimensional spectra of 66P were extracted using self-developed IDL routines. 

\subsection{UVES observations}
Observations of comet 66P were carried out in service mode with the Ultraviolet and Visual Echelle Spectrograph (UVES) mounted on the 8.2 m UT2 telescope of the European Southern Observatory. Using director discretionary time, the total 7200s of science exposure was divided into two exposures of 3600s each on July 2, 2018. We used the atmospheric dispersion corrector and the UVES standard setting DIC\#1 346 + 580, 
which roughly covers from 300 nm to 388 nm on the blue CCD and from 476 nm to 684 nm on the red mosaic CCD. We used a 0.5$\times$1.0" slit, providing a resolving power R $\sim$ 80000.

\indent The raw spectral data were reduced using the UVES Common Pipeline Library (CPL) data reduction pipeline \citep{Ballester2000}, and modified to accurately merge individual orders into a two-dimensional spectrum. Subsequently, the echelle package of the IRAF software was used to calibrate the spectra and to extract one-dimensional spectra. In turn, the cosmic rays were removed and comet spectra were rebinned and corrected for the velocity of the comet. Lastly, the continuum component, including the sunlight reflected by cometary dust grains and the telluric absorption features, was removed. 

\subsection{TRAPPIST observations}

We observed comet 66P around its perihelion (1.29 au) with the TRAPPIST-South 60-cm robotic telescope at the La Silla observatory \citep{Jehin2011}. We used the HB narrow band filters \citep{Farnham2000}, isolating the emission bands of OH[310 nm], CN[385 nm], C$_3$[405 nm], and C$_2$[515 nm] as well as emission free continuum BC[445 nm] and RC[715 nm] regions.  We also took images with broad band B, V, R$_c$, and I$_c$ Johnson-Cousin filters. Since its discovery in 1944, 66P showed a highly variable appearance during its previous perihelion passages with its visual brightness varying between 10th and 20th magnitude. We monitored the activity of this comet for two months, from May 6 to July 13, 2018. During this period, we detected the strong CN, C$_2$, and C$_3$ emissions on most of the nights while OH was only detected two times with a low signal-to-noise ratio (S/N).  NH was not detected due to the weak activity of this comet.
        
Standard procedures were used to calibrate the data by the creation of master bias, flat, and dark frames. The bias and dark subtraction, as well as the flat-field correction were done. The absolute flux calibration was made using standard stars observed during the same period \citep{Farnham2000}. We removed the sky background using the procedure developed in previous papers \citep{Opitom2015a,Opitom2015b,Opitom2016,Moulane2018}. We derived median radial brightness profiles for the gas and dust images. We removed the dust contamination from the gas radial profiles using images of the comet taken in the BC filter \citep{Farnham2000}. In order to derive the production rates, we converted the flux of the different gas species (OH, CN, C$_3,$ and C$_2$) to column densities and estimated their profiles with the Haser model \citep{Haser1957}.

For dust modeling, we used the broad band R Jonson-Cousins filters. In order to improve the S/N, the comet was imaged several times each night  using integration times in the range 60-120s. The individual images were flat-fielded and bias subtracted using standard techniques, then a median stack was obtained from the available images. The flux calibration was done using the USNO-B1.0 star catalog \citep{monet2003}. Images were calibrated in mag arcsec$^{-2}$, and then converted to solar disk units (SDUs). The USNO-B1.0 star catalog provides a photometric accuracy of 0.3 mag and 0.$''$2 of astrometric precision, which are sufficient for our modeling purposes.

\begin{table*}
        \begin{center}
                \caption{Journal of VLT and TRAPPIST Observations, and derived OH, CN, C$_2,$ and C$_3$ production rates and A($\theta$=0$^\circ$)f$\rho$ parameter of comet 66P.}
                \resizebox{\textwidth}{!}{%
                        \begin{tabular}{llcccccccccc}
                                \hline  
                                \hline
                                UT Date    & Tel & $r_h$ & $\bigtriangleup$  &  \multicolumn{4}{c}{Production rates  (10$^{24}$molecules/s)}   &  \multicolumn{4}{c}{A($\theta$=0$^\circ$)f$\rho$ (cm)} \\
                                & &(au) & (au)  & Q(OH)  & Q(CN)  & Q(C$_2$) & Q(C$_3$) & BC & RC & Rc & Ic  \\
                                \hline 
                                \hline
2018-05-06.4& TRAPPIST& 1.30&0.90&                      &5.67$\pm$0.55&4.65$\pm$0.68&                   &            &   &   &   \\
2018-05-16.4& TRAPPIST&1.29&0.90&2880$\pm$297 &8.59$\pm$0.53&8.63$\pm$0.70&2.67$\pm$0.20&82.6$\pm$6.3&108.0$\pm$7.7   &102.4$\pm$6.3   &98.6$\pm$8.6   \\
2018-05-23.4& TRAPPIST&1.29&0.90&2530$\pm$298 &8.00$\pm$0.53&8.22$\pm$0.64&2.43$\pm$0.22&75.6$\pm$5.7   &   &   &   \\
2018-05-26.4& TRAPPIST&1.29&0.90&             &7.60$\pm$0.52&7.78$\pm$0.64&2.20$\pm$0.21&70.8$\pm$6.0   &97.7$\pm$6.1   &76.7$\pm$6.9&88.0$\pm$8.6   \\
2018-06-17.4& TRAPPIST&1.34&0.91&             &6.23$\pm$0.56&3.40$\pm$0.67&1.26$\pm$0.23&   &   &46.9$\pm$6.8   &55.2$\pm$7.4   \\
2018-06-28.4& TRAPPIST&1.39&0.92&             &3.60$\pm$0.56&3.85$\pm$0.68&                   &   &   &   &   \\
2018-07-02.3&VLT/UVES&1.42 &  0.92            &&&          &   &   &   &   \\
2018-07-08.4&VLT/XSH&1.45&0.92& 891$\pm$280            &2.15$\pm$0.20&2.07$\pm$0.20& 1.92$\pm$0.20           &   &   &   &   \\
2018-07-13.4& TRAPPIST&1.48&0.92&             &1.65$\pm$0.51&1.40$\pm$0.62&             &   &   &   &   \\
2018-07-14.4& VLT/XSH&1.49&0.92& 541$\pm$230            &1.53$\pm$0.20&1.78$\pm$0.20& 1.26$\pm$0.20        &   &   &   & \\
                                \hline  
                                \hline 
                                \label{rate66P}
                        \end{tabular}}
        \end{center}
                                \tablefoot{$r_h$ and $\bigtriangleup$ are the heliocentric and geocentric distances, respectively. The A(0)f$\rho$ values are printed at 5000 km from the nucleus and they are corrected for the phase angle effect. The perihelion of 66P was on May 20, 2018 when the comet was at 1.28 au from the Sun and at 0.90 au from Earth.
                                                                }
\end{table*}

\section{Analysis and results}  

\subsection{Medium resolution spectroscopy with X-shooter}
\begin{figure}
\centering
\includegraphics[angle=00,width=\hsize]{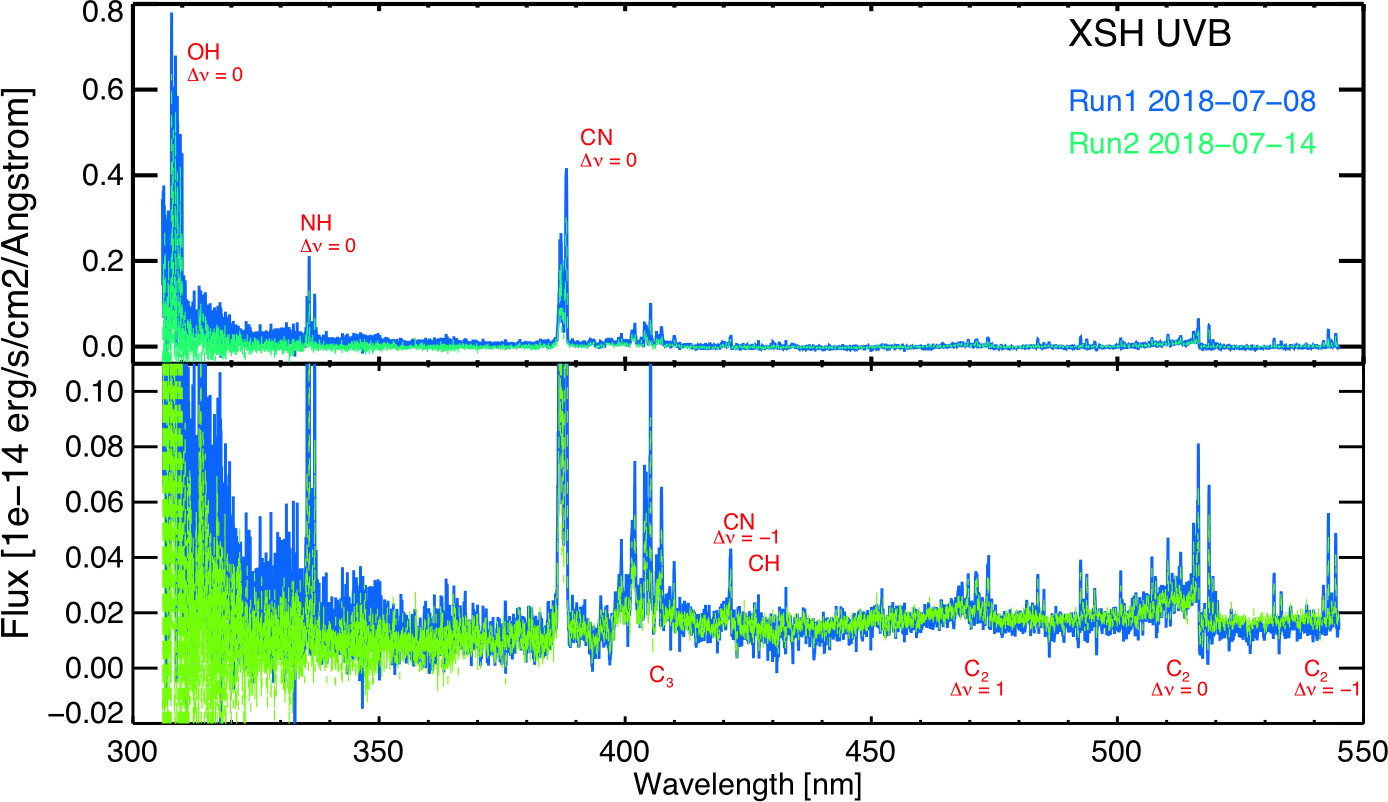}
\caption{Flux-calibrated spectra of comet 66P acquired with X-shooter on July 8 and 14. The strongest cometary emission features in the UV and optical range (OH, NH, CN, C$_2,$ and C$_3$) are marked.
}
\label{66P_emiss}
\end{figure}

X-shooter/UVB spectra of comet 66P acquired in July 2018 are shown in Fig. \ref{66P_emiss}. The first X-shooter observation was made about 1.5 months post-perihelion, the comet was still quite active as shown by the TRAPPIST observations. Several common cometary species were detected, such as OH, CN, C$_2$, C$_3,$ and NH. We derived production rates of these species using a simple Haser model \citep{Haser1957}, following the method described in \citet{Hsieh2012}. The second X-shooter observation was made a week after the first one, the comet was noticeably much fainter when further away from the Sun. Nevertheless, all the major cometary species were detected the second time. Our results of the gas production rates based on the X-shooter observations are listed in Table \ref{rate66P}, which are consistent with the values derived from the TRAPPIST data.

\begin{figure}
\centering
\includegraphics[angle=00,width=\hsize]{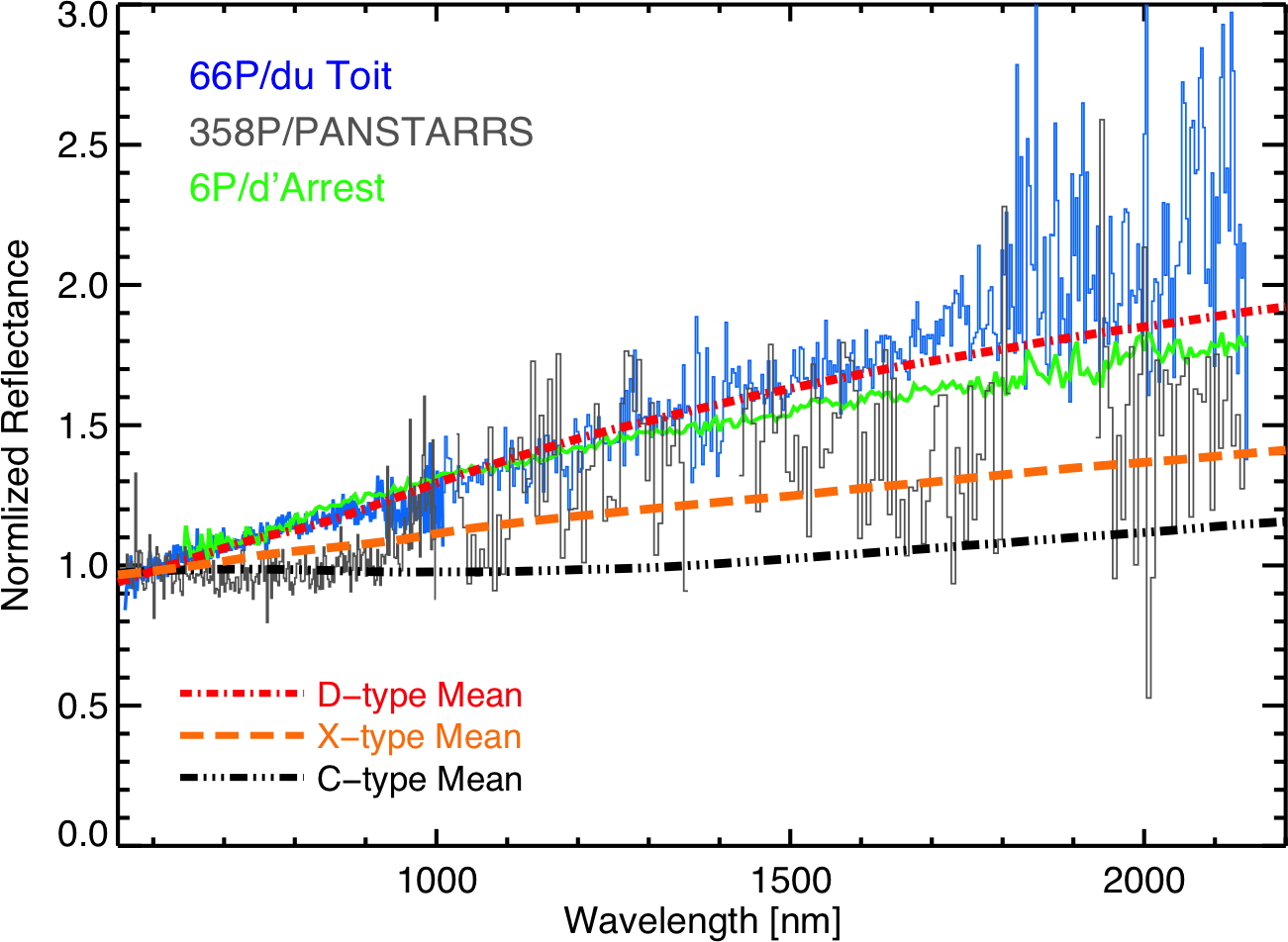}
\caption{Combined VIS and NIR reflectance spectrum of 66P is shown as the blue line. Some emission-like features in the NIR part are due to imperfect removal of the telluric absorption bands. The spectrum of the MBC, 358P, was taken with X-shooter and is from \cite{Snodgrass2017}. Spectrum of the JFC, 6P/d'Arrest, was observed with the IRTF telescope and is from \cite{yang2009}.  The comet continuum shows a red spectral slope, which is similar to the D-type (shown as the red dashed line) and is significantly deviating from the C-type (shown as the black dashed line). Three asteroid spectral classes are taken from \cite{DeMeo2009}.}
\label{66P_refl}
\end{figure}

The merged and normalized 66P spectrum in comparison to the major asteroid spectral classes \citep{DeMeo2009} is shown in Fig. \ref{66P_refl}. The relative reflectance spectrum of 66P appears featureless with no sign of the presence of water ice or hydrated minerals. The comet continuum has a reddish spectral slope, which is similar to the mean spectral slope of the D-type asteroids and significantly deviates from the mean spectral slope of the C-type asteroids. In contrast, the reflectance spectra of the majority of the known MBCs show neutral to slightly bluish slopes, which resemble the C-type spectra. 

\subsection{High resolution spectroscopy with UVES}
\begin{figure}
\centering
\includegraphics[angle=0,width=\hsize]{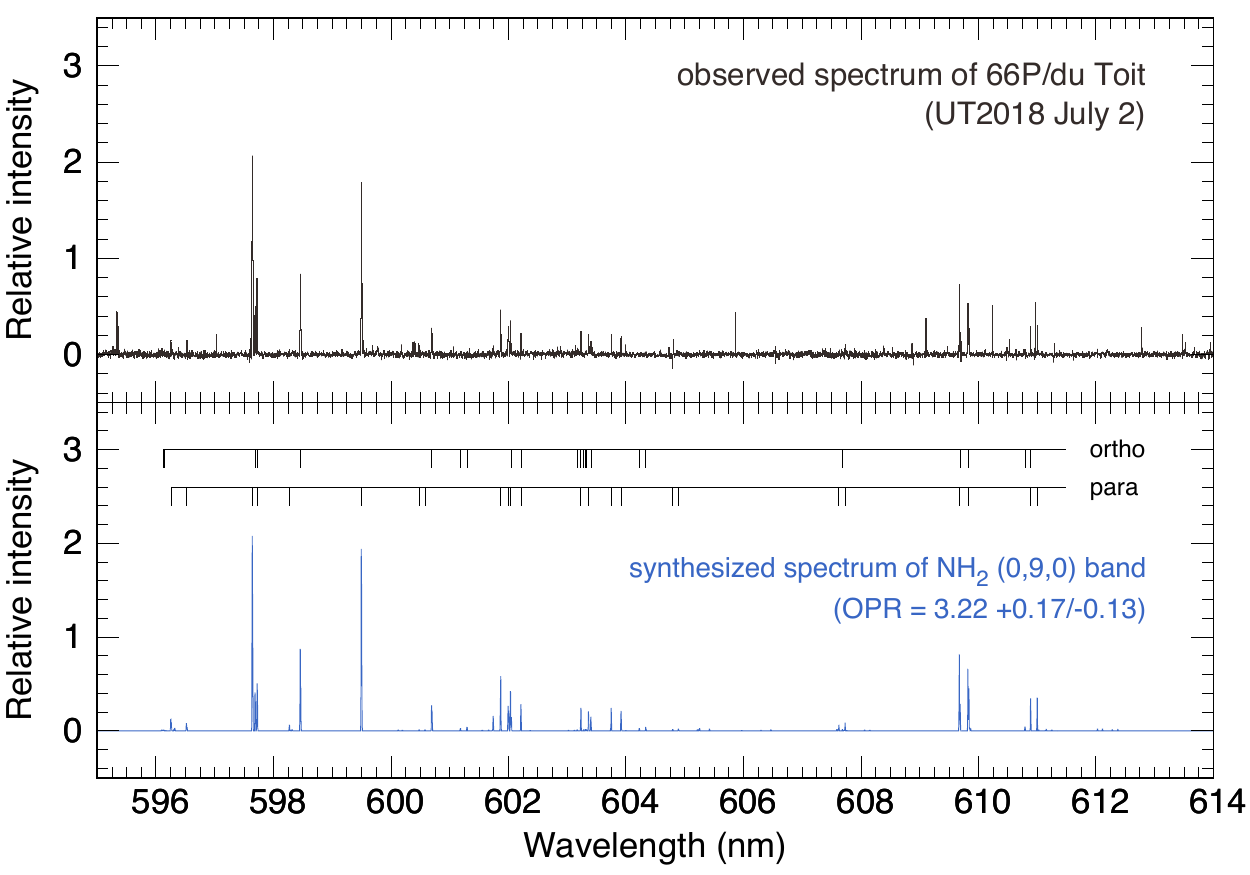}
\caption{Comparison between observed and modeled spectrum of NH$_2$.
}
\label{opr_nh2}
\end{figure}

\begin{figure}
\centering
\includegraphics[angle=-90,width=\hsize]{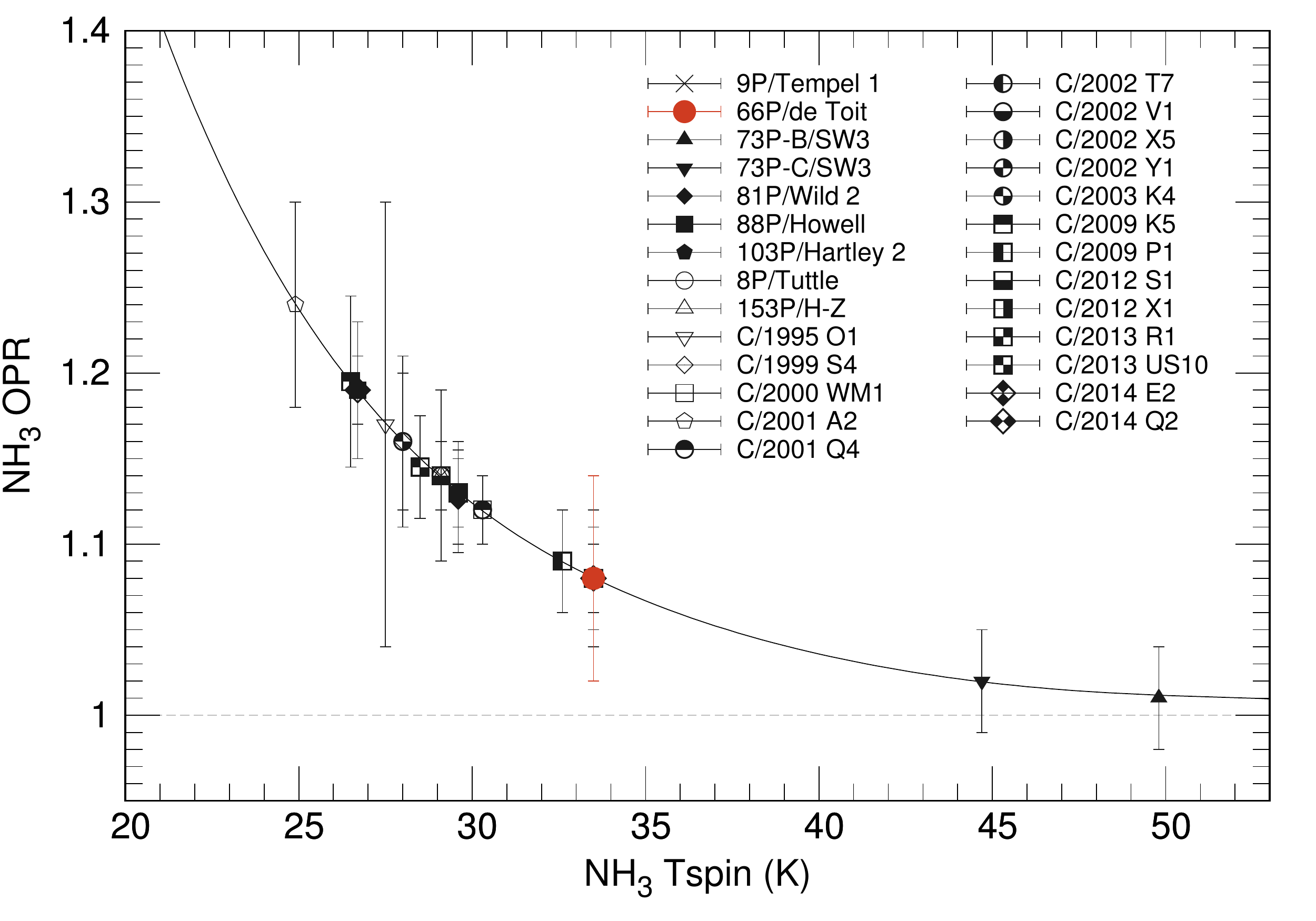}
\caption{NH$_3$ ortho-to-para ratios (OPRs) of comets converted from those of NH$_2$ and corresponding spin temperatures of NH$_3$. OPRs of other comets besides 66P are taken from  \citet{Shinnaka2016}. The horizontal dashed line indicates the nuclear-spin statistical weights ratio of ammonia (1.0).}
\label{FigVibStab}
\end{figure}

\indent Due to the close proximity of 66P at the time of the observations, we were able to measure the abundance ratio of the nuclear spin isomers of NH$_2$, namely the ortho-to-para ratio (OPR), shown in Fig. \ref{opr_nh2}.  In adopting the method described in \citet{Kawakita2001}, we further derived the OPR of NH$_3$ using the high-dispersion spectrum of NH$_2$, as seen in Table \ref{tab_opr}. Given that NH$_3$ is directly incorporated into the nucleus, the nuclear spin temperature of ammonia sets strong constraints on the formation environment of the comet. 
Our results are shown in Fig. \ref{FigVibStab}, where the NH$_3$ OPR value of 66P is comparable to those of other comets. 

We attempted to measure the nitrogen and carbon isotopic ratios from the CN violet (0,0) band following the method of \cite{Manfroid2009}. However, the observing conditions were not optimal for isotopic measurement: the comet was close to a nearly full moon and there were some cirrus clouds; moreover, the rate of outgassing of 66P was low and the total integration time for the UVES observation was not long enough to reach a desired S/N. As such, we were not able to derive either $^{12}$C/$^{13}$C or the $^{14}$N/$^{15}$N ratio.
\begin{table}[h]
\caption{Derived NH$_{2}$ and NH$_{3}$ OPRs and nuclear spin temperature}             
\label{tab_opr}      
\centering                          
\begin{tabular}{l c c c}        
\hline\hline                 
NH$_{2}$ band & NH$_{2}$ OPR & NH$_{3}$ OPR & $T_{\rm spin}$ \\    
\hline                        
   (0,7,0) & 3.18 $^{+0.17}/_{-0.13}$ & 1.09 $^{+0.09}/_{-0.07}$ & 32 $^{+12}/_{-5}$ \\      
   (0,8,0) & 3.00 $^{+0.28}/_{-0.34}$ & 1.00 $^{+0.14}/_{-0.17}$ & >21 (3-$\sigma$) \\Fig. 
   (0,9,0) & 3.22 $^{+0.21}/_{-0.29}$ & 1.11 $^{+0.11}/_{-0.15}$ & 31 $^{+27}/_{-7}$ \\
   Mean    & 3.16 $^{+0.12}/_{-0.11}$ & 1.08 $\pm$ 0.06         & 34 $^{+12}/_{-5}$ \\
\hline                                   
\end{tabular}
\end{table}

\subsection{Narrow and broad band photometry with TRAPPIST-South}

The OH, CN, C$_2$, and C$_3$ production rates as well as the  Af$\rho$ values \citep{A'Hearn1984} from TRAPPIST-South are given in Table \ref{rate66P}. The activity of 66P did not change much around perihelion (1.30-1.29 au), but it started to decrease at 1.37 au. We estimate a water production rate of about (3.24$\pm$0.17)$\times$10$^{27}$ molecules/s around perihelion, which was derived from the mean values of Q(OH) using Q(H$_2$O)=1.361$\times$r$_h^{-0.5}\times$Q(OH), given in \cite{Cochran1993}. 

 We computed the mean production rate ratios of CN/OH, C$_2$/OH, and C$_2$/CN as well as the Af$\rho$/gas ratios, such as A($\theta$=0)f$\rho$/OH and A($\theta$=0)f$\rho$/CN. Table \ref{ratios_66P} summarizes these ratios and  compares them to the typical values of comets based on the narrowband photometry survey of over 100 comets, given in \cite{Schleicher2008}. Our results show that the mixing ratio of various carbon-chain molecules of 66P are compatible with the composition of typical comets. 


\begin{table}
        \begin{center}
                \caption{Mean production rate ratios and Af$\rho$/gas ratios for comet 66P compared to  mean values of carbon-chain typical comets presented in \cite{Schleicher2008}. The A(0)f$\rho$/gas ratio has units of cm s molecules$^{-1}$.}
\begin{tabular}{lcc}
        \hline  
        \hline
        & \multicolumn{2}{c}{Log production rate ratio} \\
        \cline{2-3}
        Species &66P/du Toit   & \cite{Schleicher2008}    \\
        \hline 
        C$_2$/CN & 0.04$\pm$0.03  &  0.10   \\
        C$_2$/OH &-2.69$\pm$0.04   & -2.46    \\
        C$_3$/OH &-3.10$\pm$0.04   & -3.12   \\
        CN/OH    &-2.66$\pm$0.03   & -2.55   \\
        A($0$)f$\rho$/OH &-25.55$\pm$0.03   & -25.84$\pm$0.40$^{(a)}$  \\  
        A($0$)f$\rho$/CN &-22.87$\pm$0.03   & --  \\  
        \hline  
        \hline
        \label{ratios_66P}  
\end{tabular}
\tablefoot{$(a)$ this value is from \cite{A'Hearn1995}.}
\vspace{-1cm}
        \end{center}
\end{table}

\begin{figure}[h]
 \centering \includegraphics[scale=0.23]{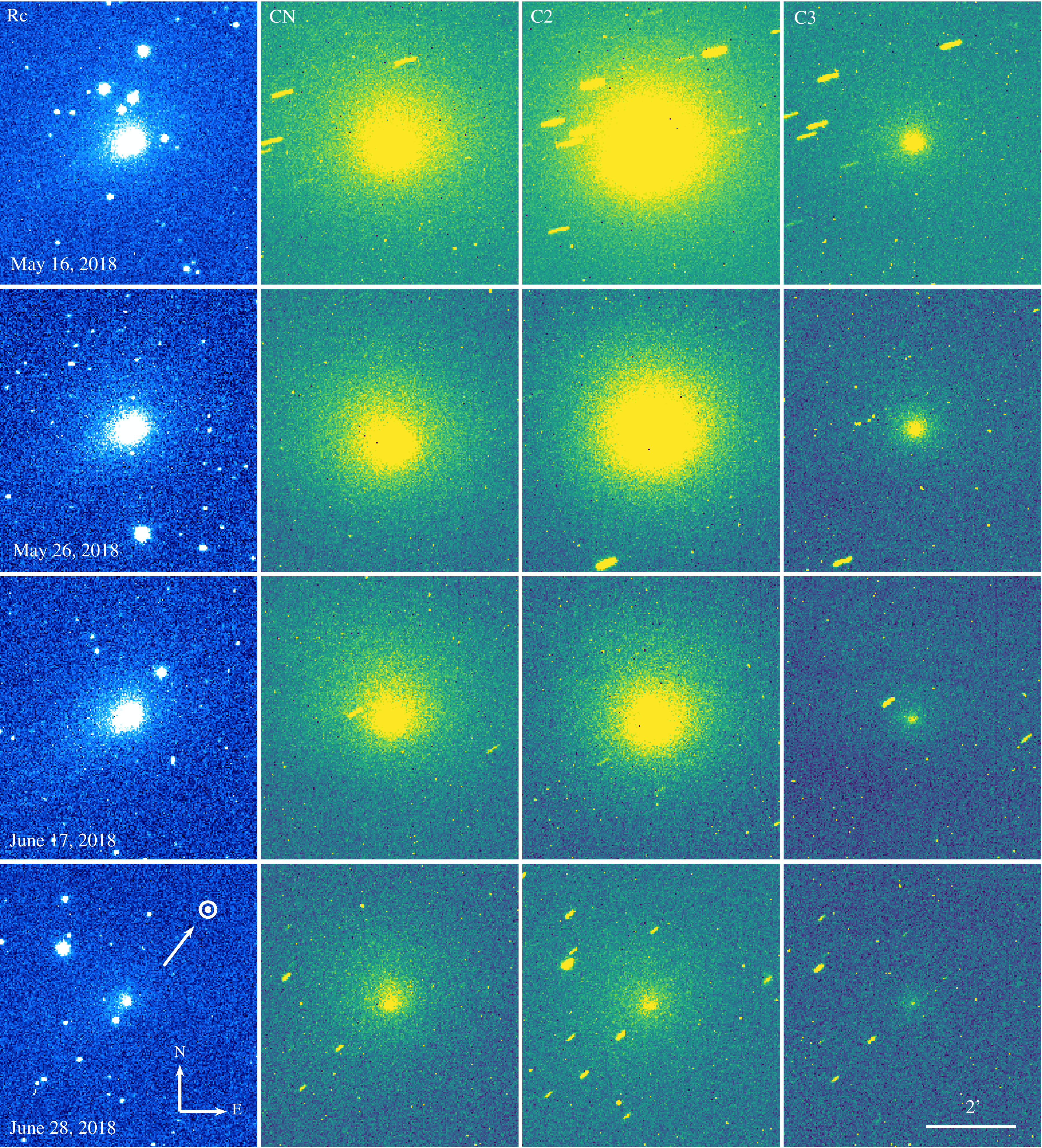}
 \caption{Evolution of 66P coma morphology in Rc, CN, C$_2,$ and C$_3$ filter images of TRAPPIST. The orientation and scale are given at the bottom of the images.}
 \label{CN_66P}
 \end{figure}

We searched for morphological features in the coma, but no jet was detected in the narrow-band or broad-band images due to the low activity of the comet. Fig. \ref{CN_66P} shows the CN coma evolution over time. It was nearly spherically symmetric throughout the monitoring window. 66P reached perihelion on May 20, 2018 when the coma was the brightest. Since then, the comet faded gradually and steadily.

\subsection{Dust environment evolution}

In order to constrain the physical properties of 66P's dust coma using the TRAPPIST observations, we 
adopted the Monte Carlo dust tail code described in \cite{moreno2012}, which has been used 
previously for several active asteroids and comets \cite[e.g.,][]{moreno2014b,moreno2016}.
The code produces synthetic images that can be compared directly with actual observations. These images are generated by adding the contribution to the brightness of each dust-like particle to mimic 
the cometary tail. 
A number of assumptions regarding the physical parameters of the dust must be made to make the problem more tractable. 
Based on in situ measurements of the dust coma of 67P from Rosetta's Optical, Spectroscopic, and Infrared Remote Imaging System (OSIRIS) and Grain Impact Analyzer and Dust Accumulator (GIADA),  we assume the density of the dust particles and the geometric albedo as $\rho=1000$ kg m$^{-3}$ and $p_{v}=0.065$, respectively \citep{fulle2016a,fornasier2015}. 
The minimum size of the particles radius, $r_{dmin}$, was set to a constant value of 1$\mu$m. This choice was motivated by results from the Micro-Imaging Dust Analysis System (MIDAS; \cite{Riedler2007}) on board Rosetta. Most of the particles of 67P were found to be hierarchical agglomerates up to a few tens of microns, which consist of micron-sized sub-units \citep{Mannel2016,Bentley2016}. Furthermore, we tested minimum values of 5 $\mu$m and 10 $\mu$m, and no significant change was observed. The maximum particle radius, $r_{dmax}(t)$, is considered as a time-dependent parameter up to decimeter sizes (\cite{rotundi2015}, \cite{fulle2016b}).
The size distribution of the particles follows a power-law function given by $n(r)\propto r^{\delta(t)}$, where  $\delta(t)$ is a time-dependent parameter 
that ranges from -4.2 to -2.0 (\cite{fulle2016a}, \cite{Ott2017}). The terminal velocity of the particles depends on the activation mechanism involved. Since the 
comet 66P showed activity during previous perihelion passages, it is most likely that the activity is driven by ice sublimation. Therefore, we assume a canonical parameterization given by $v(t,\beta)=v_{0}(t)\times\beta^{\gamma}$, where $\gamma$ is 
set to 0.5 \cite[see e.g.,][]{whipple1951,delacorte2016}. The term $v_{0}(t)$ is a time-dependent parameter. 
Besides $v_{0}(t)$, the power-law index of the particle size distribution $\delta(t)$, the maximum size of the particles, $r_{dmax}(t),$
and the dust mass loss rate, $Q_{dust}(t),$ are also time-dependent. All of these time variables are determined during the modeling process, which consists of a trial-and-error procedure, where a grid of possible combinations for the dust parameters defined is explored. In order to determine the goodness of the model during the fitting process, we computed the quantity $\chi$ for every trial following \cite{moreno2016x6}, looking for its minimum value. 

After a long set of runs using an isotropic ejection model, we find that this model does not offer a good match for the observations. Due to the poor goodness-of-fit using the isotropic model, we then considered an anisotropic ejection pattern, where the emission of the particles is characterized by active areas on the comet's surface, and the rotational state is defined by two angles \citep{sekanina1981}: the obliquity of the orbital plane to the equator, $I$, and the argument of the subsolar meridian at perihelion, $\phi$. The obliquity determines the direction of the rotation, which is prograde when $0\degr \leq I < 90\degr$ and retrograde when $90\degr < I \leq 180\degr$. When $0\degr < \phi < 180\degr$, the 
northern pole experiences sunlight at perihelion, while the southern pole receives sunlight when $180\degr < \phi < 360\degr$. In this context, we find that the anisotropic model provides a much better fit that consists of an ejection of particles coming from the northern hemisphere (ranging from $0\degr-90\degr$). The rotational parameters found are $I=(30\pm10)\degr$ and $\phi=(20\pm5)\degr$, that is, the northern hemisphere receives sunlight during the perihelion passage.

\begin{figure}[h]
   \centering
   \includegraphics[width=0.48\textwidth]{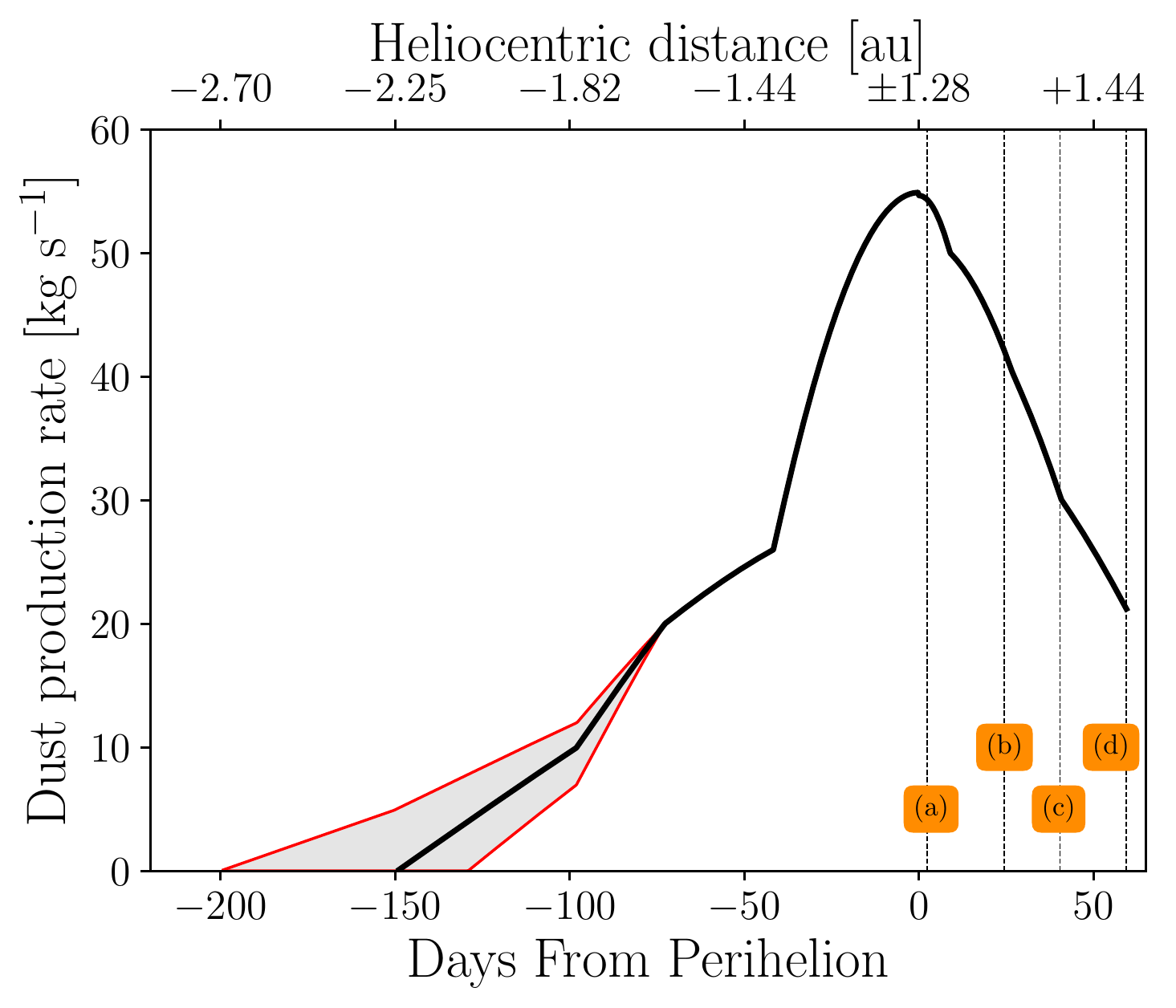}
   \caption{Dust production rate given by best-fitting model 
           as function of heliocentric distance (upper $x$-axis) and day relative to perihelion (lower $x$-axis). Dashed-vertical lines correspond to the observation dates presented in Fig. \ref{fitt} and summarized in Table \ref{table:1}. The gray shaded area corresponds to different starting dates for the activity (see text for details). }
              \label{dustprop}%
    \end{figure}

\begin{figure}[h]
   \centering
   \includegraphics[width=0.48\textwidth]{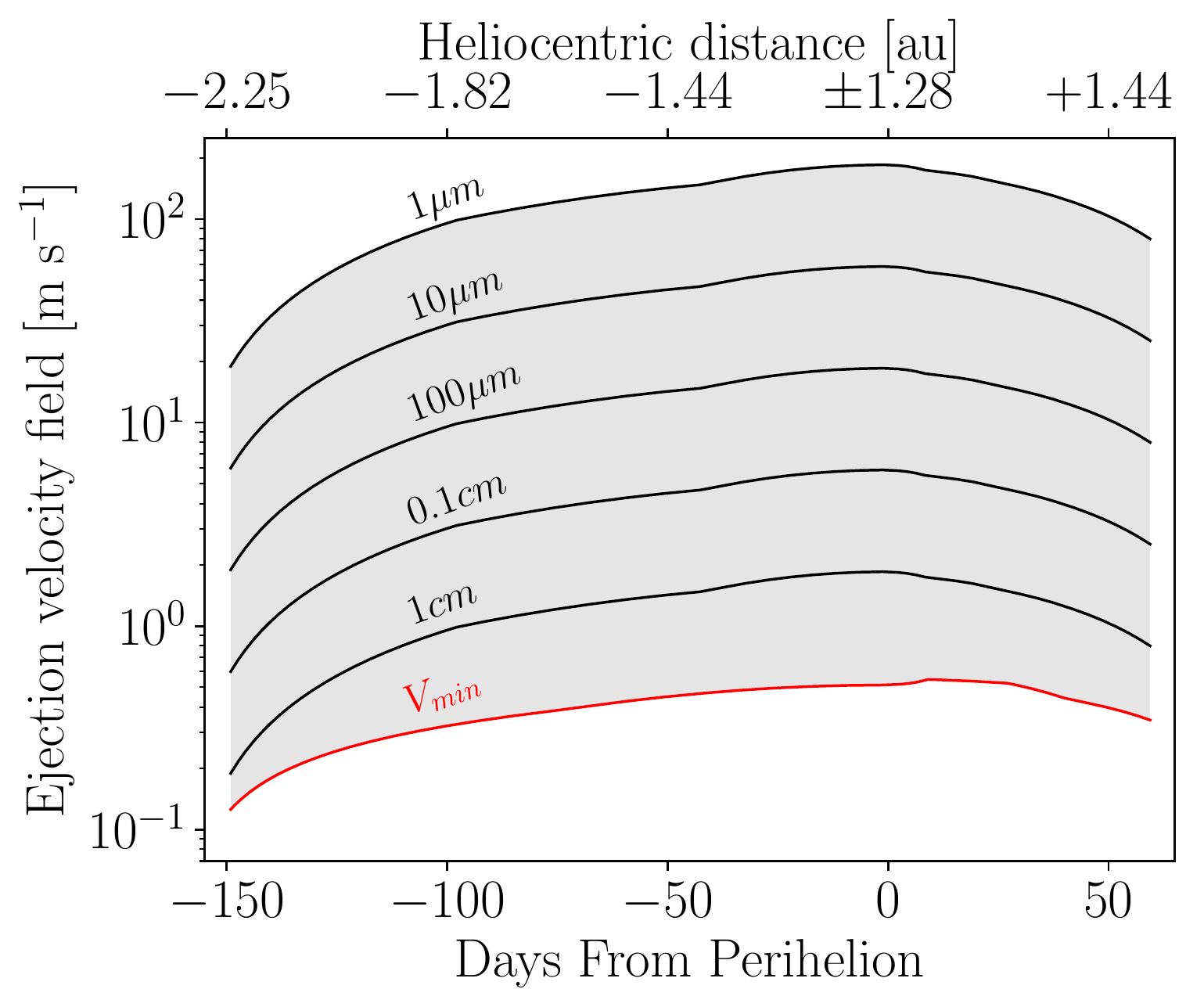}
   \caption{Ejection-velocity field of best-fitting model, as function of heliocentric distance (upper $x$-axis)
   and day relative to perihelion (lower $x$-axis). The velocities of the 1 $\mu$m, 10 $\mu$m, 100 $\mu$m, 0.1 cm, and 1 cm sized particles are shown.             
            The slowest velocity in the model is red-labeled as V$_{min}$, which corresponds to the largest particles at any moment (from 2 cm to 15 cm).}
              \label{velo}%
    \end{figure}

Overall, the best-fitting model suggests that the activity started $\sim$150 days before perihelion. However, the lack of observational information pre-perihelion,
means we can not unambiguously determine the starting date of the activity: models with starting dates between 130-200 days pre-perihelion yield 
similar results, such as $\chi(t=-200)=4.09$, $\chi(t=-150)=3.95$, and $\chi(t=-130)=4.15$. 
The dust production rate shows nearly symmetric behavior with respect to perihelion (see Fig.~\ref{dustprop}), where the peak of the 
activity is $Q_{dust}\approx~$55 kg s$^{-1}$ at perihelion distance, and the total 
dust ejected from the starting point to the last observation available on July 18 is $\sim$4.8$\times$10$^{8}$ kg. The maximum size of the particles ranges from 
2-15 cm, and the power-index of the size distribution is between -3.3 and -3.7. The ejection velocity field is displayed in Fig.~\ref{velo}. The minimum velocity 
corresponds to the largest particles at any moment, which must always be greater than the escape velocity.
To estimate the escape velocity, we adopted a nucleus radius $R_{n}=0.46$~km as reported in \cite{fernandez2015}, assuming a geometrical albedo of p$_{v}$=0.04. Regarding its density, since we are exploring the potential main belt origin of this object, it may be asteroid-like, that is, $\sim1000$~kg~m$^{-3}$ \citep{carry2012}. However, its gas and dust production rates are more similar to a typical JFC with a bulk density of $\sim550$~kg~m$^{-3}$. Using an asteroidal density or a cometary density, the corresponding escape velocity from the nucleus at a distance of 20$R_{n}$, where the gas drag vanishes, are 0.08 or 0.06 m s$^{-1}$, respectively. We considered the comet-like and asteroid-like nature of 66P and adopted the mean value of 0.07 m s$^{-1}$. As shown in Figure \ref{velo}, the maximum speed is $\sim$200~m~s$^{-1}$ at perihelion distance ($\sim$1.28 au) and achieved by the smallest particles in the model. The comparisons, found here, between the maximum speed and the results derived by other authors using Monte Carlo models or other dust models show that similar terminal velocities were found for micron-sized particles ejected by other comets. \cite[see e.g.,][]{pozuelos2018,moreno201767p,Agarwal2007}

\begin{figure}
   \centering
   \includegraphics[width=0.48\textwidth]{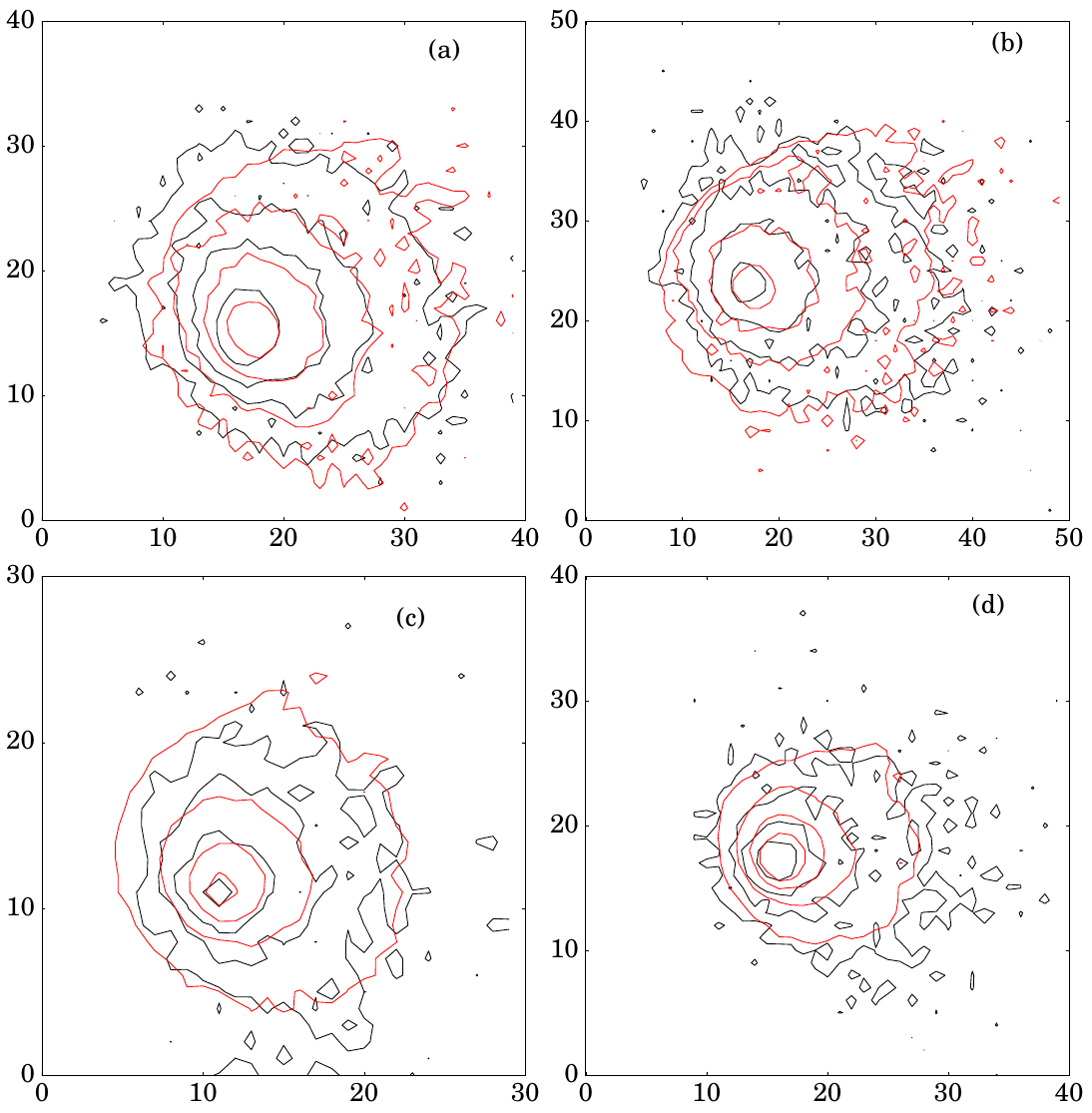}
   \caption{Comparison of four observed (see Table \ref{table:1}) and modeled images. In all cases, the black contours correspond to 
            observations and the red ones to the best-fitting model. The $y$ and $x$ axes are given in units of pixels.
            The plot is oriented so that north is up and east is to the left. 
                 }
              \label{fitt}%
    \end{figure}

In Fig. \ref{fitt}, four images selected from the observational data set are compared with the corresponding synthetic images, which were generated by the best-fitting model. Due to a large number of time-dependent variables (e.g., $v_{0}(t)$, $\delta(t)$, $r_{dmax}(t)$, and $Q_{dust}(t)$ ) in the model, it might be possible to find an alternative set of parameter values that could also fit the observational data. Further constraints on the model can be implemented by obtaining future observations that cover a significant orbital arc and include both pre- and post-perihelion, which was the case for comet 41P \citep[see e.g.,][]{pozuelos2018}.

\begin{table*}
\caption{Log of  observations used for  dust modeling.}             
\label{table:1}      
\centering 
\begin{tabular}{cc c c c c c}     
\hline\hline       
\noalign{\smallskip}                     
Date  & Days to & Resolution$^{(1)}$ & Dimension $^{(2)}$ & Phase & $A(\theta=0^{\circ})f\rho$ $^{(3)}$  & $A(\theta=0^{\circ})f\rho^{\prime}$ $^{(4)}$ \\
(UT) & perihelion & (km pixel$^{-1}$) & (pixels$^{2}$) & Angle ($\degr$) & (cm) & (cm)  \\

\noalign{\smallskip} 
\hline  
\noalign{\smallskip}                  
(a) 2018-05-23.4  & 2.4  & 832.7 & 40 & 51.3 & 86.8$\pm$4.3 & 79.7$\pm$6.3  \\
(b) 2018-06-14.4 & 24.4 & 847.6 & 50 & 49.3 & 62.5$\pm$3.1 & 66.4$\pm$4.1   \\ 
(c) 2018-06-28.4 & 38.4 & 856.0 & 30 & 46.7 & 37.8$\pm$1.8 & 35.0$\pm$3.5 \\ 
(d) 2018-07-19.4 & 59.4 & 856.0 & 40 & 40.5 & 29.8$\pm$1.5 & 31.9$\pm$2.7  \\ 

\noalign{\smallskip}
\hline
\noalign{\smallskip}

\end{tabular}
 \tablefoot{
 $^{(1)}$ Resolution of the images in Fig. \ref{fitt}. $^{(2)}$ Dimensions of the images in Fig. \ref{fitt}. $^{(3)}$ Corresponding to the observations at $\rho=5000$ km. $^{(4)}$ Corresponding to the synthetic images at $\rho=5000$ km.}
            
\end{table*}

\section{Dynamical evolution}

\cite{fernandez2015} defined a likely dynamical path to determine the degree of orbital stability for the near-Earth JFCs. They find a strong correlation between the orbital stability and a set of critical parameters, which are  the indices $f_{q}$ and $f_{a}$, the capture-time, $t_{cap}$, and the closest approach to Jupiter, $d_{min}$. In short, the $f_{q}$ index evaluates the time spent under the gravitational influence of Jupiter during the last 10$^{4}$ yr, that is, the fraction of time during the last 10$^{4}$ yr that the comet (or any of its clones) moves along an orbit with $q > 2.5$ au, or reaches heliocentric distances $r_h > $100 au. The index $f_{a}$ refers to the time during which the comet (or any of its clones) orbits with $a > 7.37$ au, that is, the comet is no longer controlled by Jupiter. Additionally, the capture-time is the time in the past at which the mean perihelion, $\bar{q}(t)$, increased by one au with respect to the initial value at the discovery time. This parameter describes the time spent by a comet in Earth's vicinity. We refer the reader to \cite{fernandez2015} for a full mathematical description of the indices $f_{q}$, $f_{a}$, and $t_{cap}$.

Unstable near-Earth JFCs have their $f_{q}$ and $f_{a} \gg$ 0 and a small fraction of near-Earth JFCs with very stable orbits have nearly zero $f_{q}$ and $f_{a}$ as well as very large $t_{cap}$ ($ > 5\times10^{4}$ yr) and $\bar{d}_{min}>0.3$ au \citep{fernandez2015}. In this context, the comet 66P was found to have $f_{q}= f_{a}= 0$, $t_{cap}> 5\times10^{4}$ yr, and $\bar{d}_{min}>1.00$ au (2.87 Jupiter's hill radii). Thus, 66P is considered highly asteroidal and may have its origin in the asteroid belt.

After the last perihelion passage, the orbit of 66P has the best possible accuracy. We revisited and computed the parameters $f_{q}$, $f_{a}$, $t_{cap,}$ and $\bar{d}_{min}$ to verify the results of \cite{fernandez2015} using the latest orbital elements. We used numerical integrations in the heliocentric frame, starting from January 1$^{}$, 2019 and integrated backward up to $10^{5}$ yr. We adopted the numerical package MERCURY \citep{chambers1999}, with   
the Bulirsch-Stoer algorithm, which offers a high accuracy integration, but it is slow. 
To perform the statistical study, we generated 200 clones of the nominal orbit of 66P, according to the associated 6$\times$6 covariance matrix \citep{cherni1998}. 

\begin{figure}
   \centering
   \includegraphics[width=0.5\textwidth]{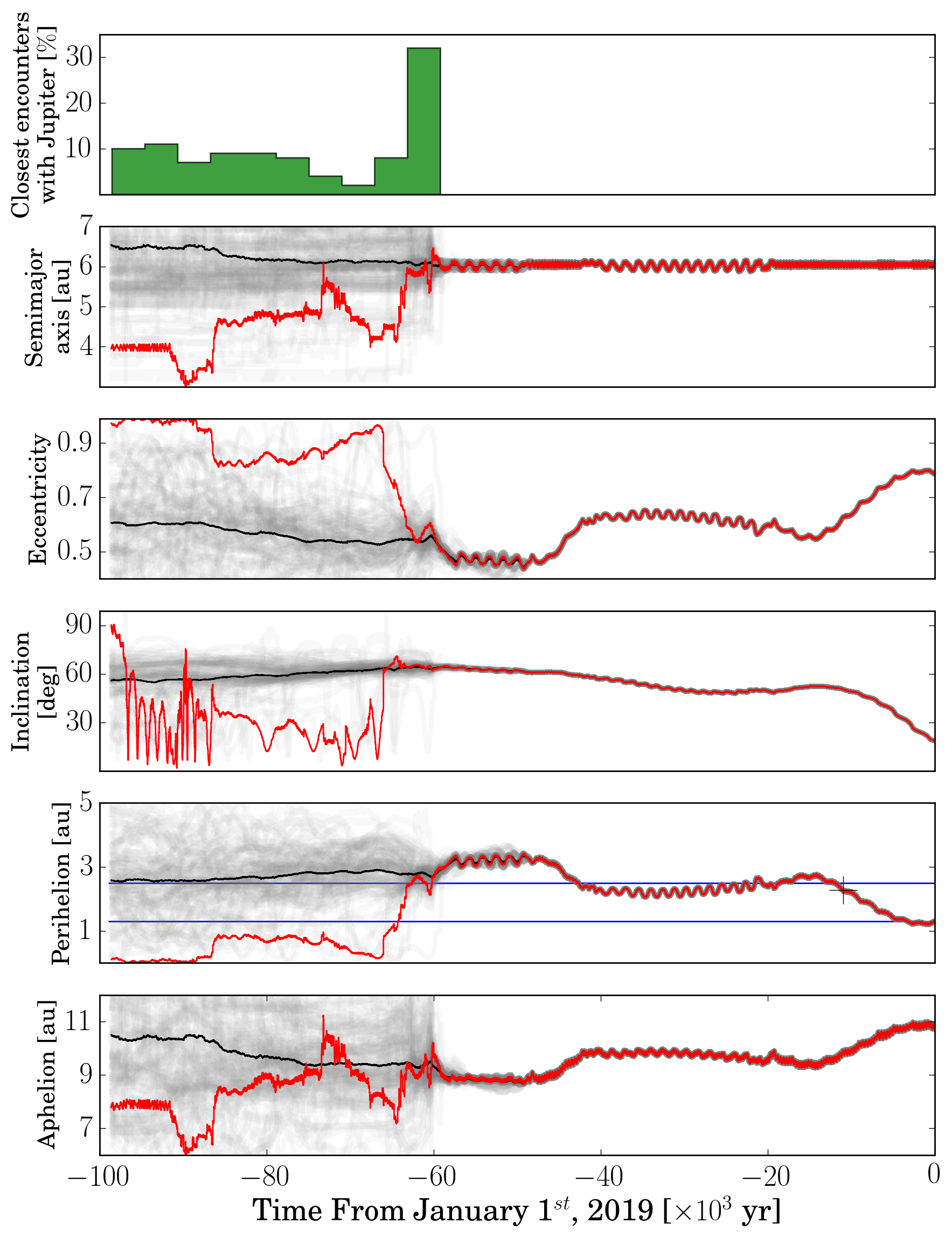}
   \caption{Orbital evolution of 66P and its 200 clones for $10^{5}$ yr backward in time from January 1, 2019. From the top to the bottom: 
            the closest approaches with Jupiter, semi-major axis, eccentricity, inclination, perihelion, and aphelion distance. In all cases the red lines 
            correspond to the evolution of the nominal comet 66P, the black lines are the mean values of the whole set of 200 clones plus the nominal comet, and the 
            gray lines show the individual evolution for each clone. In the perihelion panel the black cross indicates the time of the capture, $t_{cap}$, and the 
            two horizontal blue bars correspond to $q=1.3$ and $q=2.5$ au.
            The initial orbital elements were taken from the JPL Small-Body Data Browser.}
              \label{evol_clones}%
    \end{figure}

We ran our simulations the second time to examine the influence of the non-gravitational forces; in this case, the covariance matrix includes two extra terms, which are the radial and transverse acceleration, respectively. Both sets of the orbital parameters and the covariance matrix of the orbit for 66P are published together in the NASA/JPL small-body browser. In order to ensure the highest accuracy possible, we integrated every clone 
independently, that is, we performed 200 simulations with one clone each. The initial time-step was set to five days and
the computed orbital evolution was stored every year for each clone. The Sun, the eight planets, and Pluto were included in the simulation.
Because of the weak activity of 66P, the simulations with and without non-gravitational forces yielded identical results. Hence, in the subsequent 
analysis we only show the results of the pure gravitational model. 
The dynamical history of 66P and its clones are shown in Fig \ref{evol_clones}. We notice that the orbital evolution of 66P and of its clones are extremely compact 
during a period of 6.0$\times10^{4}$ yr, with no significant divergence. During this period, the comet and the clones are safe from 
close encounters with Jupiter, with minimum distances greater than 1 au. At the time of $\sim$-6.0$\times10^{4}$ yr, the orbits begin to diverge and display chaotic behavior. This effect matches
with the closest encounter with Jupiter of $\bar{d}_{min}\sim0.2$ au (0.57 Jupiter's hill radii). About 30$\%$ of the clones have their closest encounter at that time, and the rest of them 
during the period -6 to -10 $\times10^{4}$ yr. We find
$f_{q}$=0.0, $f_{a}$=0.0, which are consistent with the values given in \cite{fernandez2015}. However, the capture time, $t_{cap}\sim 10^{4}$ yr, is about an order of magnitude shorter. According to the formal definition presented in \cite{fernandez2015}, 66P can no longer be categorized as highly asteroidal because of the shorter $t_{cap}$. Instead, 66P is only moderately asteroidal and therefore cannot be considered as a NEMBC candidate.

\begin{figure}
   \centering
  \includegraphics[width=0.5\textwidth]{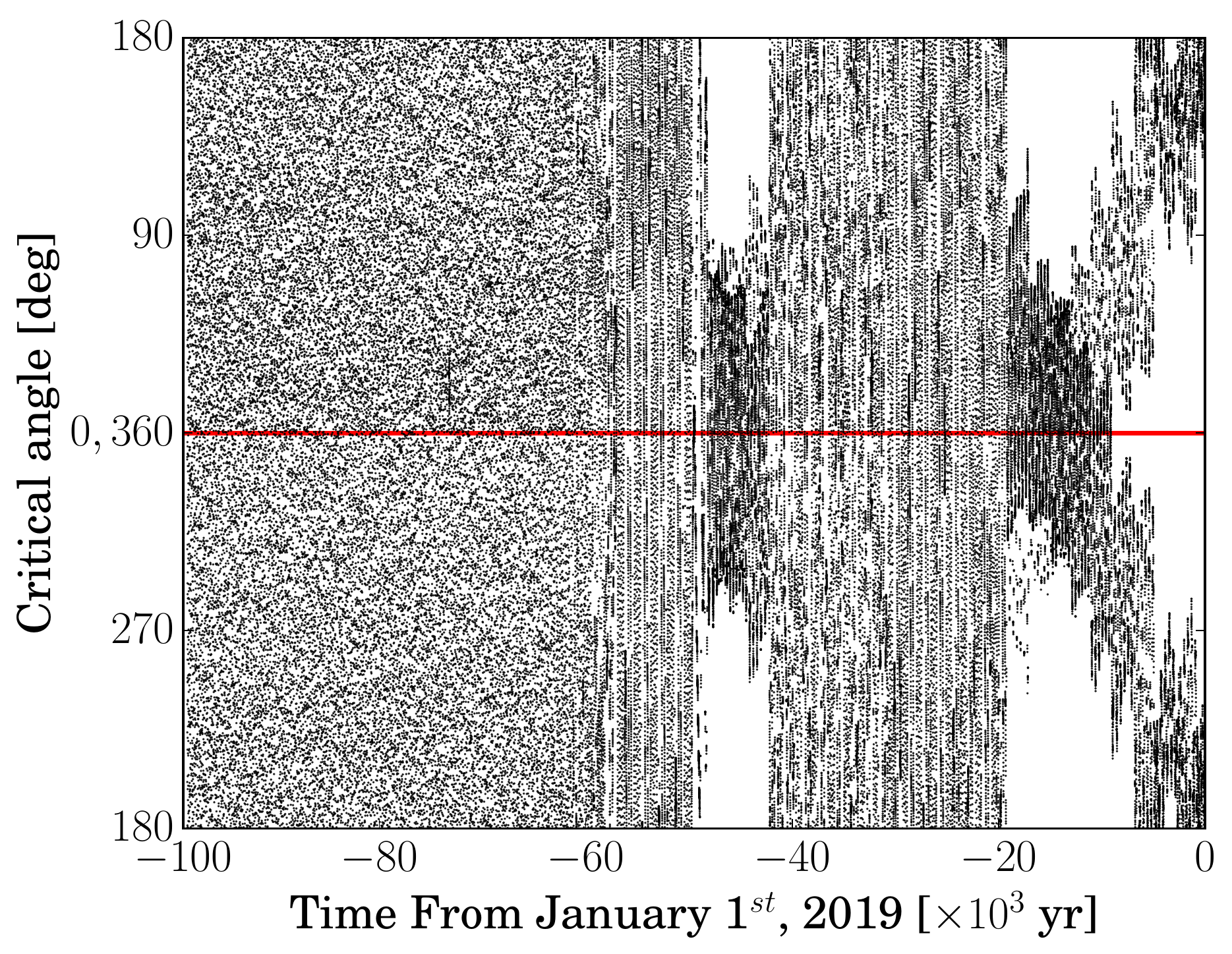}
   \caption{Evolution of critical angle of 66P defined as $\sigma_{c}=(p+q)\lambda_{J}-p\lambda_{c}-q\bar{\omega}$, where 
            $\textbar p+q\textbar : \textbar p\textbar $  corresponds to MMR with Jupiter, $\lambda_{J}$ is mean 
            longitude of Jupiter, $\lambda_{c}$  mean longitude of 66P, and $\bar{\omega}$ is longitude of perihelion.
            }
              \label{critical_angles}%
    \end{figure}

On the other hand, our simulations show that $q$ oscillates between 1.3 au and 2.5 au for $\sim4.0\times10^{4}$ yr. This pattern is similar to the behavior of most near-Earth asteroids, whose orbits are stable for $10^{4}$ yr or longer, with perihelion distances confined to $q<2.5$ au and semi-major axes to $a<7.37$ au. Many of these asteroids are trapped in mean motion resonances (MMRs) with Jupiter \citep{fernandez2014}. In addition, we noticed that the semimajor axis of 66P oscillates around 
a quasi-constant value of 6 au, so we calculated the strength for all possible resonances located near this value following \cite{gallardo2006}. We find that 66P 
is trapped in 4:5 MMR with Jupiter from 0 to -60$\times10^{3}$ yr, with different states of the critical angle, $\sigma_{c,}$ as shown in Fig. \ref{critical_angles}. 
During the time that 66P is trapped in the MMR, the evolution of $\sigma_{c}$ is complex, which alternates between periods of libration and periods of circulation. 
From 0 to -5000 yr, $\sigma_{c}$ librates around 180$^{\circ}$,  with a large semi-amplitude of 90$^{\circ}$. From -5000 to -10$\times10^{3}$ yr it gently evolves toward 0$^{\circ}$, where it remains until -20$\times10^{3}$ yr. From  
-20$\times10^{3}$ yr to -45$\times10^{3}$ yr $\sigma_{c}$ circulates, and back again to librate around 0$^{\circ}$ from -45$\times10^{3}$ yr to -50$\times10^{3}$ yr. Finally,
from -50$\times10^{3}$ yr to -60$\times10^{3}$ yr it circulates again, before showing chaotic behavior, presumably due to the close encounter with Jupiter at this time. It is interesting to note that during the period that $\sigma_{c}$ circulates, there is a coupling of the orbital parameters 
as shown in Fig. \ref{evol_clones}. Although $i$ shows very little cyclic variations, it may be coupled with $q$, indicating the action of the
Kozai mechanism, where the parameter $H=\sqrt{1-e^{2}}~ cos(i)$ remains constant. We computed the evolution 
of $H$, and we obtained a non-constant value that varied from 0.60 to 0.35 during the time that 66P was trapped in the MMR from 0 to -60$\times10^{3}$ yr. As such, we conclude that the Kozai mechanism is not responsible for the coupling of the orbital parameters.

\section{Discussion}
Based on our UVES observations of 66P, we derived an NH$_3$ OPR of 1.08 $\pm$ 0.06 and a spin temperature of 34K, which are consistent with the values observed in others JFCs \citep{Shinnaka2016}. Traditionally, the OPRs exhibited by gaseous species observed in cometary comae have been used to constrain the formation temperature of comets, assuming that OPRs have been unchanged in cometary nuclei since the formation of the molecules about 4.6 billion years ago. However, there are several problems with this view. As shown in Fig. \ref{FigVibStab}, most comets have NH$_3$ OPRs between 1.1 - 1.2, corresponding to a spin temperature of $\sim$ 30K. This temperature is significantly higher than 10K as suggested by the theoretical studies on $^{15}$N-fractionation in ammonia \citep{Shinnaka2016}. Recently, laboratory studies have found that OPR of H$_2$O can be modified and re-equilibrated via interactions with other molecules or with the solid matrix \citep[and references therein]{Hama2016}. The new findings of the laboratory studies suggest that it is incorrect to assume the OPRs of cometary species are primordial and the OPRs can not be used to derive the formation temperature of certain molecules, such as T$_{spin}$. More detailed discussions on OPRs are present in \citep{Shinnaka2016}. Similar to water, the OPR of ammonia can be modified in the coma by electron recombination, we, therefore, can no longer use the derived NH$_3$ OPR to constrain the nucleus formation temperature and, in turn, to verify the origin of 66P.

\begin{figure}
\centering \includegraphics[scale=3.0]{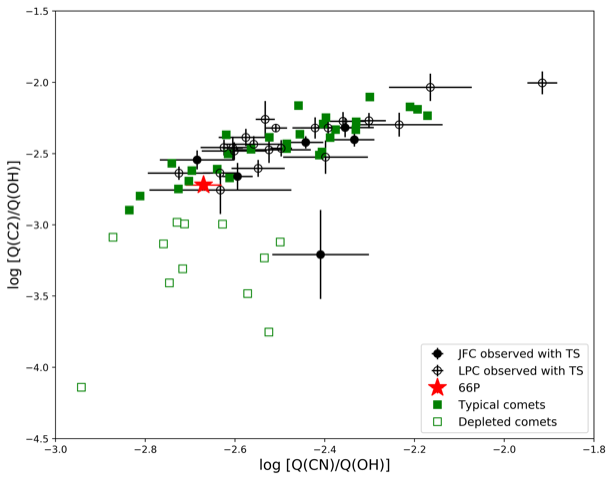}
\caption{Logarithm of  ratio of C$_2$ to OH production rates as function of logarithm of  ratio of CN to OH for 66P (red star) compared to JFCs (filled circles) and long-period comets (crosses) observed with TRAPPIST between 2010 and 2016 \citep{Opitom2016thesis}, and also with typical (green filled squares) and carbon-chain depleted (green opened squares) comets given in \cite{A'Hearn1995}.
}
\label{Ahearn}
\end{figure}

Although 66P is a weakly active comet, thanks to its close distance to the Earth at the time of observations, several gas species were detected both with X-shooter/VLT and TRAPPIST. The relative abundances, such as the Q(CN)/Q(OH) ratio versus the Q(C$_2$)/Q(OH) ratio of 66P as well as those of JFCs and Oort cloud comets are shown in Fig. \ref{Ahearn}. The Q(CN)/Q(OH) and Q(C$_2$)/Q(OH) of 66P are slightly lower than other JFCs; however, its relative abundance ratios are within the normal values of over 100 typical comets studied in \citep{Schleicher2008, A'Hearn1995}. We note that 66P was observed only within a narrow time window. Nevertheless, the TRAPPIST observations of 66P were made at typical heliocentric distances for other JFC observations, so it is reasonable to compare our results with other JFC measurements. 

\cite{A'Hearn1995} noted that for a typical comet all gaseous species vary at a similar rate with the heliocentric distance, which can be described as Q(gas)$ \propto r_h^{-2.7}$. Given that the activity of MBCs is likely to be driven by water ice sublimation, we scaled the upper limits of water and CN production rates of the known MBCs to $r_h$=1.29 au, where most observations were made for 66P. The scaled values as well as the original upper limits are listed in Table \ref{MBCs_66P}. Q(H$_2$O) of 66P is about an order of magnitude higher than those of known MBCs. One caveat of this comparison is that most upper limits of Q(H$_2$O) of MBCs are scaled from Q(CN), assuming a cometary ratio Q(CN)/Q(H$_2$O) $\sim$ 0.001. MBCs are closer to the Sun than typical comets, therefore CN could be depleted in MBCs \citep{Prialnik2009}. However, for 176P and 358P, the Q(H$_2$O) rates were derived by observing the H$_2$O and OH lines directly \citep{de2012upper,ORourke2013}, which are comparable to the limits derived using Q(CN). 

\begin{table*}
        \begin{center}
                \caption{Upper limits of CN and H$_2$O production rates and dust production rates of MBCs \citep{Snodgrass2017bb} compared to 66P.}
                        \begin{tabular}{lcccccccl}
                                \hline  
                                \hline
                                Objects & Q(H$_2$O)$_s^a$ & Q(CN)$_s^a$ & $\frac{dM}{dt}_s^b$ & r$_h$ & Q(H$_2$O) & Q(CN) &  $\frac{dM}{dt}$ &References \\
                                      & ($10^{26}$ molec/s)     & ($10^{24}$molec/s)   & (kg/s) & (au) & ( $10^{26}$molec/s)     & ($10^{24}$molec/s)   & (kg/s) & \\
                                \hline 
                                133P   & 0.1   &0.1 & 7.2  & 2.64 & 0.02 & 0.01 & 1.4 &\cite{licandro2011}\\
                                176P  &  2.6    &-  & 0.5  & 2.58 & 0.40 & -  & 0.1 & \cite{de2012upper}\\
                                259P    & 1.3    &0.4 & - & 1.86 & 0.50 & 0.14  & - & \cite{jewitt2009main}\\
                                288P    & 6.1   &2.6 & 2.3 & 2.52 & 1.00 & 0.42  & 0.5 & \cite{hsieh2012discovery}\\
                                313P   & 3.2    &1.6 & 1.7 & 2.41 & 0.60 & 0.18  & 0.4 & \cite{Jewitt2015}\\
                             324P  &  7.1   &1.9 & 1.1  & 2.66 & 1.00 & 0.30  & 0.2 & \cite{hsieh2012observational}\\
                             358P  & 2.7  &0.8  & 4.3 & 2.42 & 0.50 & 0.15 & 1.0  & \cite{Hsieh2013}\\
                   P/2013 R3  & 1.9  &1.2  & $<$3.5  & 2.23 & 0.43 & 0.12 & $<$1.0$^c$ & \citep{jewitt2014disintegrating}\\
                                \hline
                                66P      &27.1    &7.5   &55.0& 1.29 & 27.1 & 7.5  &55.0& This Work \\
                                \hline 
                                        \label{MBCs_66P}
                        \end{tabular}
                \end{center}
                \tablefoot{\tablefoottext{a}{Scaled water and CN production rates to $r_h$=1.29 au, using Q(gas)$ \propto r_h^{-2.7}$ given by \cite{A'Hearn1995}}\\
                \tablefoottext{b}{Scaled dust production rates to $r_h$=1.29 au, using the proxy relationship: $\frac{dM}{dt}  \propto r_h^{-2.3}$ , given by \cite{A'Hearn1995}}\\
              \tablefoottext{c}{The empirical limit to the mass loss, from \cite{Jewitt2017}}
}
        \end{table*}

The optical colors and spectra of known MBCs are mostly similar to those of the C-class asteroids \citep{DeMeo2009}. Only one MBC, 358P/PANSTARRS, has been observed in the NIR and the comet appears redder in the NIR as shown in Fig.  \ref{66P_refl}. However, 358P was very faint at the time of the observations and its NIR spectrum is rather noisy and heavily affected by the telluric absorptions. Our X-shooter observations show that the D-type like spectrum of 66P is significantly different from those of MBCs and is more similar to the spectra of active JFCs, such as 6P/d'Arrest. We would like to point out that the MBC spectra as well as the spectrum of 66P were obtained when these objects were active. Therefore, the spectra do not necessarily reflect the intrinsic composition of the nuclei because the nuclei were buried inside the dust comae along the light-of-sight. Besides composition, the spectral slope can also be affected by the size distribution of the dust coma. Although we can not exclusively conclude that the intrinsic composition of 66P differs from those of the MBCs, our observations suggest that the composition and/or the particle size distribution of the coma of 66P are more similar to those of the JFCs than the MBCs.  

Moreover, our dust models find remarkable differences between 66P and the MBCs, the latter in general show much lower values for both dust production rates (0.2-1.4 kg s$^{-1}$) and ejection velocities (0.5-2.0 m s$^{-1}$ for particles of 100 $\mu$m) \citep[see e.g.,][]{hesieh2009a,jewitt2014,pozuelos2015,agarwal2016}.  
In comparison, the JFCs have typical dust productions rates of 40-250 kg s$^{-1}$ and ejection velocities for particles of 100 $\mu$m ranging from 5-10 m s$^{-1}$
\cite[see e.g.,][]{pozuelos2014,pozuelos201481p,moreno201767p}. The previous statistical study of 85 comets found the average variation of the dust production rate can be expressed as A($\theta$)f$\rho \propto r_h^{-2.3}$ \citep{A'Hearn1995}. When scaling to $r_h$=1.29 au, the dust production rates of MBCs would be in the range of 1-10 kg s$^{-1}$,  as listed in the Table \ref{MBCs_66P}, which are at least one order of magnitude smaller than the dust production rate of 66P.


Using the latest orbital elements, our dynamical simulation confirmed that the orbit of 66P is stable with $f_{q}$ and $f_{a}$ equal to 0. However, our model found a much shorter $t_{cap}$. Therefore 66P is no longer highly asteroidal and can not be considered as an NEMBC candidate. The discrepancy between our results and those of \cite{fernandez2015} may either be caused by the quality of the orbital elements or by the number of clones and how they are generated. \cite{fernandez2015} used a simple Gaussian distribution with a standard deviation given by the nominal uncertainties to generate 50 clones. In contrast, we used a more robust method based on the covariance matrix and we generated 200 clones in this study. On the other hand, a detailed exploration of its dynamical evolution shows that 66P does share some features with near-Earth asteroids. 

For other NEMBC candidates identified in \citet{fernandez2015}, \cite{fernandez2017} studied the anomalous comet 249P/Linear and found that its activity lasts only $\sim$20 days around perihelion with a peak mass loss rate of 145 kg s$^{-1}$ that is much higher than the dust production rates of MBCs. The spectrum of 249P was found to be similar to B-type asteroids, a trait that is also shared with some MBCs. Although the dynamical simulations show some similarities for 249P and 66P, the physical properties of these two objects are very different. 

\section{Summary and conclusion}
We performed detailed physical studies of the near-Earth JFC: 66P, our main results are:\\
\indent Firstly, based on the UVES/VLT observations, a suit of NH$_2$ lines were detected. The OPR of NH$_3$ using the high-dispersion spectrum of NH$_2$ was found to be 1.08$\pm$ 0.06 and its T$_{spin}$=34K$^{+12}_{-5}$, which are comparable to the values of other normal JFCs.\\ 
\indent Secondly, based on the X-shooter/VLT and the TRAPPIST observations, common cometary gaseous species such as, OH, CN, C$_2,$ and C$_3$ were detected. The relative abundances of 66P is consistent with those of typical JFCs. The reflectance spectrum of 66P closely resembles the mean spectrum of the D-type asteroids, which is much redder than the spectra of the known MBCs.\\ 
\indent Thirdly, dust models using the TRAPPIST observations obtained the peak mass loss rate of 55 kg s$^{-1}$ that is about an order of magnitude larger than the mass loss rates of the MBCs when scaled to the same heliocentric distance.\\
\indent Lastly, dynamical simulations using the latest orbital elements of 66P found that it is no longer highly asteroidal but moderately asteroidal due to the shorter capture time.\\ 

Considering all the available observations as well as the results of the dust model and the dynamical model, we conclude that 66P is much more similar to typical JFCs than MBCs and, therefore, it is unlikely to have originated from the asteroid main belt and it is not related to MBCs. 

\begin{acknowledgements}
We would like to thank the anonymous referee and the editor, Emmanuel Lellouch, for their careful review and constructive suggestions. Based on observations collected at the European Organisation for Astronomical Research in the Southern Hemisphere under ESO program 2101.C-5033(A) and (B). TRAPPIST-South is funded by the Belgian Fund for Scientific Research (Fond National de la Recherche Scientifique, FNRS) under the grant FRFC 2.5.594.09.F. EJ and DH are Belgian FNRS Senior Research Associates. Simulations in this paper made use of the REBOUND code which can be downloaded freely at http://github.com/hannorein/rebound.
\end{acknowledgements}


\end{document}